\documentstyle[epsfig]{mn}
 
\def\be{\begin{equation}} 
\def\ee{\end{equation}}

\def\HI{\hbox{H~$\scriptstyle\rm I\ $}}

\def\gsim{\lower.5ex\hbox{\gtsima}} 
\def\lsim{\lower.5ex\hbox{\ltsima}} \def\gtsima{$\; \buildrel > \over 
\sim \;$} \def\ltsima{$\; \buildrel < \over \sim \;$} \def\prosima{$\; 
\buildrel \propto \over \sim \;$} \def\gsim{\lower.5ex\hbox{\gtsima}} 
\def\lsim{\lower.5ex\hbox{\ltsima}} 
\def\simgt{\lower.5ex\hbox{\gtsima}} 
\def\simlt{\lower.5ex\hbox{\ltsima}} 
\def\simpr{\lower.5ex\hbox{\prosima}}   
  
\def\gtsima{$\; \buildrel > \over \sim \;$} 
\def\ltsima{$\; \buildrel < \over \sim \;$} 
\def\gsim{\lower.5ex\hbox{\gtsima}} 
\def\lsim{\lower.5ex\hbox{\ltsima}} 
\def\simgt{\lower.5ex\hbox{\gtsima}} 
\def\simlt{\lower.5ex\hbox{\ltsima}} 
\def\simpr{\lower.5ex\hbox{\prosima}}

\def\Msun{M_{\odot}} 
\def\Zsun{Z_{\odot}}

\def\E3{{\cal E}_{\rm g}^{III}}

\title{High redshift Ly$\alpha$ emitters: clues on the Milky Way infancy}
\author[Salvadori, Dayal \& Ferrara]{Stefania Salvadori$^{1}$\thanks{E-mail:salvadori@astro.rug.nl}, Pratika Dayal$^{2}$ \& Andrea Ferrara $^{3}$\\ 
$^{{1}}$ Kapteyn Astronomical Institute, Landlaven 12, 9747 AD Groningen, The Netherlands\\ 
$^{2}$  SISSA/International School for Advanced Studies, Via Beirut 2-4, 34014 Trieste, Italy\\
$^{3}$ Scuola Normale Superiore, Piazza dei Cavalieri 7, 56126 Pisa, Italy}

\begin{document} 
\date{} 
\pagerange{\pageref{firstpage}--\pageref{lastpage}} \pubyear{} 
\maketitle 

\label{firstpage} 
\begin{abstract}
With the aim of determining if Milky Way (MW) progenitors could be identified 
as high redshift Lyman Alpha Emitters (LAEs) we have derived the intrinsic 
properties of $z\approx 5.7$ MW progenitors, which are then used to compute 
their observed Ly$\alpha$ luminosity, $L_\alpha$, and equivalent width, EW. MW progenitors visible 
as LAEs are selected according to the canonical observational criterion, 
$L_\alpha>10^{42}$~erg~s$^{-1}$ and $EW>20$ \AA. Progenitors of MW-like 
galaxies have $L_{\alpha} =10^{39-43.25}$~erg~s$^{-1}$, making some of 
them visible as LAEs. In any single MW merger tree realization, typically 
only 1 (out of $\approx 50$) progenitor meets the LAE selection criterion, but the 
probability to have {\it at least} one LAE  is very high, $P=68\%$. The 
identified LAE stars have ages, $t_* \approx  150-400$~Myr at $z\approx 5.7$ 
with the exception of five small progenitors with  $t_* < 5$~Myr and large 
$EW=60-130$~\AA. LAE MW progenitors provide $> 10\%$ of the halo very 
metal-poor stars [Fe/H]$<-2$, thus establishing a potentially fruitful link 
between high-$z$ galaxies and the Local Universe.
\end{abstract} 


\section{Introduction} 
\label{intro}
Lyman Alpha Emitters (LAEs) are galaxies identified by means of a very strong 
Ly$\alpha$ line (1216 \AA). Advances in instrument sensitivity and specific 
spectral signatures (strength, width and continuum break bluewards of the line) 
have enabled the confirmation of hundreds of LAEs in a wide redshift range, at 
$z\approx 2.25$ (Nilsson et al. 2008), $z\approx3$ (Cowie \& Hu 1998; Steidel 
et al. 2000; Matsuda et al. 2005; Venemans et al. 2007; Ouchi et al. 2008), 
$z\approx 4.5$ (Finkelstein et al. 2007), $z\approx5.7$ (Malhotra et al. 2005; 
Shimasaku et al. 2006) and $z\approx 6.6$ (Taniguchi et al. 2005; Kashikawa 
set al. 2006). 

LAEs have by now been used extensively as probes of both the ionization state 
of the intergalactic medium (IGM) and probes of high redshift galaxy evolution 
(Santos 2004; Dayal, Ferrara \& Gallerani 2008; Nagamine et al. 2008; Dayal et 
al. 2009; Dayal, Ferrara \& Saro 2010; Dayal, Maselli \& Ferrara 2010). 
However, in spite of the growing data sets, there has been no effort to 
establish a link between the properties of these early galaxies to observations 
of the local Universe, {\it in primis} the Milky Way (MW). Our aim in this work 
is to investigate the possible connection between the Galactic building blocks 
and LAEs at a time when the Universe was $\approx 1$ Gyr old. This will allow 
us to answer to questions as: are the progenitors of MW-like galaxies visible 
as LAEs at high redshifts? How can we discriminate amongst LAEs which are 
possible MW progenitors and those that are not? What are the physical 
properties of these Galactic building blocks?

To this end, we adopt a novel approach of coupling the semi-analytical code 
{\tt GAMETE} (Salvadori, Schneider \& Ferrara 2007; Salvadori, Ferrara \& 
Schneider 2008, Salvadori \& Ferrara 2009), which traces the hierarchical 
build-up of the Galaxy, successfully reproducing most of the observed MW 
and dwarf satellite properties at $z=0$, to a previously developed LAE model 
(Dayal, Ferrara \& Gallerani 2008; Dayal et al. 2009; Dayal, Ferrara \& Saro 
2010), that reproduces a number of important observational data sets accumulated 
for high-z LAEs.
\section{Obtaining the MW progenitors}
\label{gamete}
We start by summarizing the main features of {\tt GAMETE}, which is used to 
build-up 80 possible hierarchical merger histories and to derive the 
properties of the MW progenitors at $z=5.7$\footnote {We specifically choose 
$z=5.7$ for all our calculations as it represents the highest redshift for 
which a statistically significant sample of confirmed LAEs is available.}. 
First, the possible hierarchical merger histories of a MW-size dark matter 
(DM) halo are reconstructed up to $z=20$ via a Monte Carlo algorithm based 
on the extended Press-Schechter theory (see Salvadori, Schneider \& Ferrara 
2007 for more details). The evolution of gas and stars is then followed along 
each hierarchical tree by assuming that: 
(a) the initial gas content of DM haloes is equal to the universal cosmological 
value $(\Omega_b/\Omega_m)M_h$, where $M_h$ is the DM halo mass, 
$\Omega_b$ ($\Omega_m)$ is the baryonic (DM) density parameter; 
(b) at any redshift, there exists a minimum halo mass to form stars, 
$M_{sf}(z)$, whose evolution accounts for the suppression of star formation 
(SF) in progressively more massive objects due to radiative feedback effects 
(see Fig.~1 of Salvadori \& Ferrara 2009); (c) the gradual accretion of cold 
gas, $M_c$, into newly virializing haloes is regulated by a numerically 
calibrated infall rate (Kere\v{s} et~al. 2005); (d) the SF rate, $\dot M_*$, 
is proportional to the mass of cold gas inside each galaxy, 
$\dot M_*=\epsilon_* M_c/t_{ff}$, where $\epsilon_*$ is the SF efficiency 
and $t_{ff}$ the halo free fall time; (e) in halos with a virial temperature, 
$T_{vir}<10^4$~K (minihalos), the SF efficiency is reduced as 
$\epsilon = \epsilon_*[1+(T_{vir}/2\times 10^4{\rm K})^{-3}]^{-1}$ due to 
ineffective cooling by H$_2$ molecules. The chemical enrichment of gas, both in 
the proto-Galactic halos and in the MW environment is followed simultaneously 
by taking into account the mass-dependent stellar evolutionary timescales and 
the effects of mechanical feedback due to supernova (SN) energy deposition 
(see Salvadori, Ferrara \& Schneider 2008 for more details).
\begin{figure}
\psfig{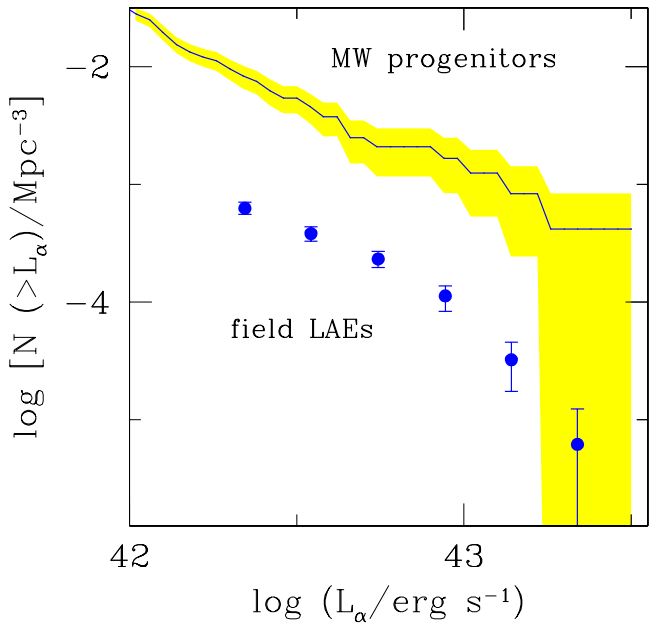}
              \caption{Cumulative Ly$\alpha$ luminosity function at 
       redshift $z \approx 5.7$. The solid line shows the mean Ly$\alpha$ 
       LF of the MW progenitors averaged over 80 realizations of the merger 
       tree; the dashed area represents the $\pm 1 \sigma$ spread among 
       different realizations. For comparison only we show (points) the 
       field LAE LF (Shimasaku et~al. 2006).}
  \label{lyalf}
\end{figure}

The two free parameters of the model (star formation and wind efficiencies) 
are calibrated by reproducing the global properties of the MW (stellar/gas 
mass and metallicity) and the Metallicity Distribution Function (MDF) of 
Galactic halo stars (Salvadori, Schneider \& Ferrara 2007, Salvadori, 
Ferrara \& Schneider 2008); $M_{sf}(z)$ is fixed by matching 
the observed iron-luminosity relation for dwarf spheroidal galaxies 
(Salvadori \& Ferrara 2009). They are assumed to be the same for all 
progenitors in the hierarchical tree. 
\section{Identifying LAEs}
\label{laes}
By using {\tt GAMETE} we obtain the total halo/stellar/gas masses ($M_h$, $M_*$, $M_g$), the instantaneous SF rate ($\dot M_*$), the mass weighted stellar metallicity ($Z_*$) and the mass-weighted stellar age ($t_*$) of each MW progenitor, in each of the 80 realizations considered. These outputs are used to calculate the total intrinsic Ly$\alpha$ ($L_\alpha^{int}$) and continuum luminosity ($L_c^{int}$) which include both the contribution from stellar sources and from the cooling of collisionally excited neutral hydrogen (\HI) in the interstellar medium (ISM) (Dayal, Ferrara \& Saro 2010).

The intrinsic Ly$\alpha$ luminosity can be translated into the observed luminosity such that $L_\alpha = L_\alpha^{int} f_\alpha T_\alpha$, while the observed 
continuum luminosity, $L_c$ is expressed as $L_c = L_c^{int} f_c$. Here, $f_\alpha$ ($f_c$) are the fractions of Ly$\alpha$ (continuum) photons escaping the galaxy, undamped by the ISM dust and $T_\alpha$ is the fraction of the Ly$\alpha$ luminosity that is transmitted through the IGM, undamped by \HI \footnote{The continuum band (1250-1500 \AA) is chosen so as to be unaffected by \HI absorption.}.

The main features of the model used to calculate $f_c, f_\alpha$ and $T_\alpha$ 
are: (a) for each MW progenitor the dust enrichment is derived by using its 
intrinsic properties ($\dot M_*$, $t_*$, $M_g$) and assuming Type II supernovae 
(SNII) to be the primary dust factories. The dust mass, $M_d$, is calculated 
including dust production due to SNII (each SN produces $0.5 \Msun$ of dust), 
dust destruction with an efficiency of about 40\% in the region shocked to 
speeds $\geq 100~{\rm km \, s^{-1}}$ by SNII shocks, assimilation of a 
homogeneous mixture of dust and gas into subsequent SF (astration), and 
ejection of a homogeneous mixture of gas and dust from the galaxy due to SNII.
(b) Following the calculations presented in Dayal, Maselli \& Ferrara (2010) 
for high-z LAEs, the dust distribution radius, $r_d$, is taken to scale with the 
gas distribution scale, $r_g$, such that $r_d \sim 0.5 r_g$ \footnote{the gas 
distribution radius is calculated as $r_g = 4.5\lambda r_{200}$, where the spin 
parameter, $\lambda=0.05$ (Ferrara, Pettini \& Shchekinov 2000) and $r_{200}$ 
is the virial radius.} (c) $f_c$ is calculated assuming 
a slab-like dust distribution and we use $f_\alpha = 1.3 f_c$, as inferred 
for LAEs at $z \approx 6$. (d) $T_\alpha$ is calculated using the mean 
photoionization rate predicted by the Early Reionization Model (ERM, 
reionization ends at $z \approx 7$) of Gallerani et al. (2008), according 
to which the neutral hydrogen fraction $\chi_{HI} = 7 \times 10^{-5}$ at 
$z \approx 5.7$. Complete details of these calculations can be found in Dayal, 
Ferrara \& Saro (2010) and Dayal, Maselli \& Ferrara (2010).
\begin{figure}
  \centerline{\psfig{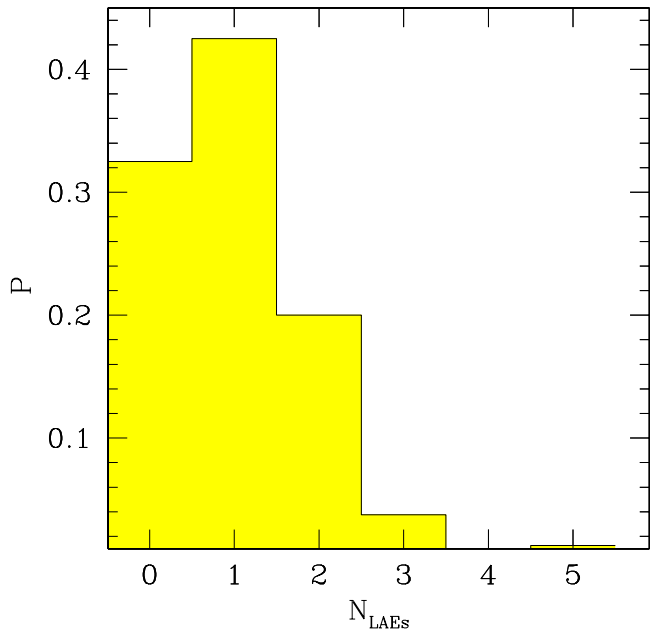}}
  \caption{Probability of finding a number, $N_{LAEs}$, of LAEs in a single
        MW realization, averaged over 80 hierarchical merger histories at
        $z \approx 5.7$.}
  \label{freq}
\end{figure}

Progenitors are then identified as LAEs based on the currently used 
observational criterion: $L_\alpha \geq 10^{42} \, {\rm erg \, s^{-1}}$ 
and the observed equivalent width, $L_\alpha/L_c \geq 20$~\AA. We use a 
comoving volume of the Milky Way environment ($30 {\rm Mpc}^3$) to 
calculate the number density of the progenitors visible as LAEs and the 
observed LAE luminosity function (LF), shown in Fig. \ref{lyalf}.
\section{Results}
\label{results}
We start by comparing the number density of field LAEs observed at 
$z\approx 5.7$ with the average LF of the Milky Way progenitors at the same 
epoch (Fig.~\ref{lyalf}). The number density of MW progenitors decreases with 
increasing Ly$\alpha$ luminosity, reflecting the higher abundance of the least 
massive/luminous objects in $\Lambda$CDM models. The MW progenitors cover the 
entire range of observed Ly$\alpha$ luminosities, $L_{\alpha}=10^{42-43.25} 
{\rm erg~s^{-1}}$. We then conclude that {\it among the LAEs observed at 
$z\approx 5.7$ there are progenitors of MW-like galaxies}. For any given 
$L_\alpha$ the number density of MW progenitors is higher than the observed 
value because the MW environment is a high-density, biased region. As discusses 
in the final Section, uncertainties on the treatment of dust may also play a role.  

In Fig.~\ref{freq} we show the probability distribution function (PDF), $P$, of 
finding a number $N_{LAEs}$ of LAEs in any given MW realization. The PDF has a 
maximum ($P=0.42$) at $N_{LAEs}=1$; while in any single MW realization, there 
are $\approx 50$ star-forming progenitors with stellar masses $M_* \gsim 10^7\Msun$, 
on average only one of them would be visible as a LAE. We note that $P=0.32$ for 
$N_{LAEs}=0$, while it rapidly declines ($P<0.2$) for $N_{LAEs}>1$. We conclude 
that the MW progenitors that would be observable as LAEs at $z\approx 5.7$ are rare 
($\approx 1/50$), but the probability to have {\it at least} one LAE in any 
MW hierarchical merger history is very high, $P=68\%$. 
\begin{figure*}
\centerline{\psfig{figure=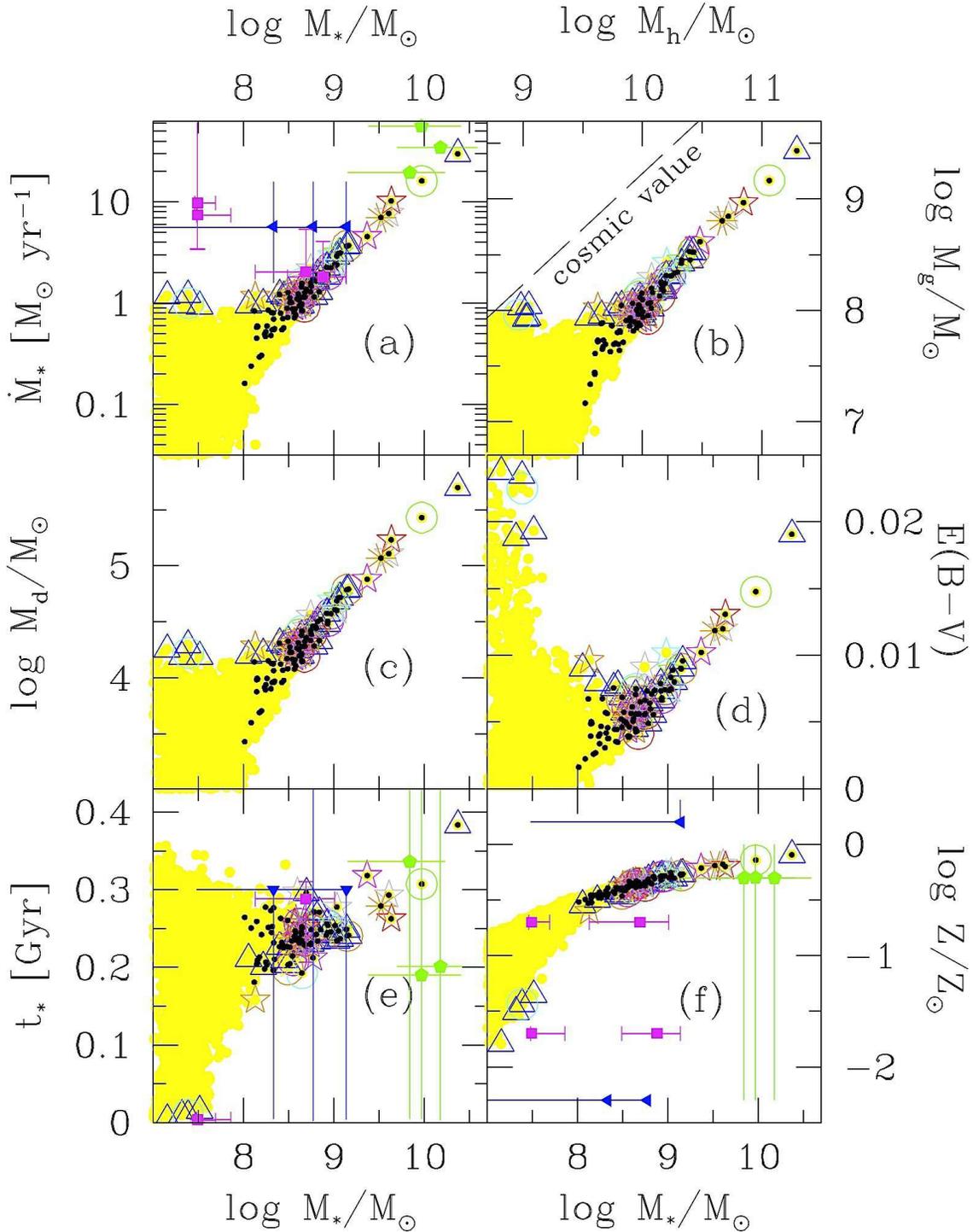,width=15.0cm,angle=0}}
\caption{Physical properties of $z \approx 5.7$ MW progenitors in 80 
different hierarchical merger histories. We show: (i) all the progenitors 
(yellow circles), (ii) the major branches of each hierarchical tree (black 
filled points) and (iii) the progenitors identified as LAEs (colored open 
symbols). LAEs pertaining to the same (different) realizations are shown 
with the identical (different) colored symbols (see the text and Fig. 
\ref{freq}). Triangles are used for those realizations in which there is 
only one LAE. As a function of the total stellar mass $M_*$ the various 
panels show: (a) the instantaneous star formation rate, $\dot M_*$; (c) 
the dust mass, $M_d$; (d) the color excess $E(B-V)$; (e) the average 
stellar age, $t_*$; (f) the average stellar metallicity $Z$. Panel (b) 
shows the relation between the halo and gas mass, with the cosmic value 
$(\Omega_b/\Omega_m) M_h$ pointed out by the dashed line. Points with 
errorbars are the observational LAE data collected by Ono et~al. 2010 
(magenta squares, 1 LAE, four different models), Pirzkal et~al. 2007 
(blue triangles, 3 LAEs) and Lai et~al. 2007 (green circles, 3 LAEs).}
\label{prop}
\end{figure*}

Let us now consider the physical properties of the building blocks of the MW. 
$\dot M_*$ represents the dominant physical factor to determine whether a 
progenitor would be visible as a LAE, since it governs  the intrinsic 
Ly$\alpha$/continuum luminosity,  the dust enrichment (and hence absorption)
and $T_\alpha$, as both the size of the ionized region around each source 
and the \HI ionization fraction inside it scale with $\dot M_* $ (Dayal, 
Ferrara \& Gallerani 2008). This implies the existence of a SF rate threshold 
for MW progenitors to be visible as LAEs which is 
$\dot M^{min}_* \approx 0.9\Msun$yr$^{-1}$ (panel (a) of Fig.~\ref{prop}). 
Since $\dot M_* \propto M_g$ (see Sec.~2), such a lower limit can be translated 
into a threshold gas mass: $M^{min}_g\approx 8\times 10^7\Msun$ (panel (b)). 
In turn, the gas content of a (proto-) galaxy is predominantly determined by 
the assembling history of its halo and the effects of  SN (mechanical) 
feedback. While the most massive MW progenitors ($M_h > 10^{10}\Msun$) display 
a tight $M_g-M_h$ correlation, the least massive ones are highly scattered, 
reflecting the strong dispersion in the formation epoch/history of recently 
assembled halos (panel (b)). As a consequence, {\it LAEs typically correspond 
to the most massive progenitors of the hierarchical tree}, i.e. the major 
branches (black points in the panels). In particular,  we find that {\it all} 
haloes with $M_h \geq 10^{10}\Msun$ (40 haloes) are LAEs. At decreasing $M_h$, 
instead, the progenitors can be visible as LAEs only by virtue of a high gas 
mass content or extremely young ages; there are 5 such objects, with 
$M_h \approx 10^9 \Msun$, as seen from panel (b).

By comparing the panels (b) and (c) of Fig.~\ref{prop}, we can see that $M_d$ 
of the Galactic building blocks closely tracks $M_g$. Since the gas mass 
content of the $z\approx 5.7$ MW progenitors is reduced by $\approx 1$ order 
of magnitude with respect to the initial cosmic value, due to gas (and dust) 
loss in galactic winds, their resulting dust mass is relatively small: 
$M_d\approx 10^{4-5.7}\Msun$. As a consequence all the progenitor
galaxies have a color excess E(B-V)$<0.025$. In panel (d) we see that as expected,  
E(B-V) increases with $M_*$, i.e. more massive galaxies are redder. However 
the trend is inverted for $M_*< 10^8 \Msun$; even though the dust masses in 
these low mass objects ($M < 10^9 \Msun$) is $M_d \lsim 10^{4.2}\Msun$, due 
to their small virial radius, both the gas distribution scale, and hence the 
dust distribution scale are very small (see Sec. \ref{laes}). The concentration 
of the dust in a small area leads to a large dust attenuation and hence, a 
large value of the color excess.

MW progenitors visible as LAEs are generally intermediate age objects, 
$t_* \approx 150 -400$ Myr, as seen from panel (e); the largest progenitors 
tend to be the oldest ones, an expected feature of standard hierarchical 
structure formation scenarios. However, the ages show a large scatter, 
especially at decreasing $M_*$, reflecting the great variety of assembling 
(and SF) histories of recently formed halos. Interestingly the five 
$M_h\approx 10^9\Msun$ newly formed ($z < 6$) progenitors visible as LAEs 
have a very young stellar population, $t_*\leq 5$~Myr. The high $\dot M_*$ 
induced by the large mass reservoir and the copious Ly$\alpha$ production 
from these young stars makes them detectable.

In panel (f) we can see that these newly virializing galaxies are
metal-poor objects, with an average stellar metallicity  $Z\approx 0.016-0.044 
\Zsun$. As these galaxies host a single and extremely young stellar population, 
such low $Z$ value reflects the metallicity of the MW environment at their 
formation epoch, $z\approx 5.7$ (see the middle panel in Fig.~1 of Salvadori, 
Ferrara \& Schneider 2008). 
The more massive MW progenitors visible as LAEs, instead, are more metal rich, 
$Z\approx 0.3-1 \Zsun$; their intermediate stellar populations form during a 
long period (see Fig.~4) and from a gas that is progressively enriched by 
different stellar generations.

The fact that the physical properties of the LAEs progenitors obtained with our
model are consistent with those inferred from the observations of field LAEs 
(Fig.~3) is a notable success of our model.
\begin{figure}
  \centerline{\psfig{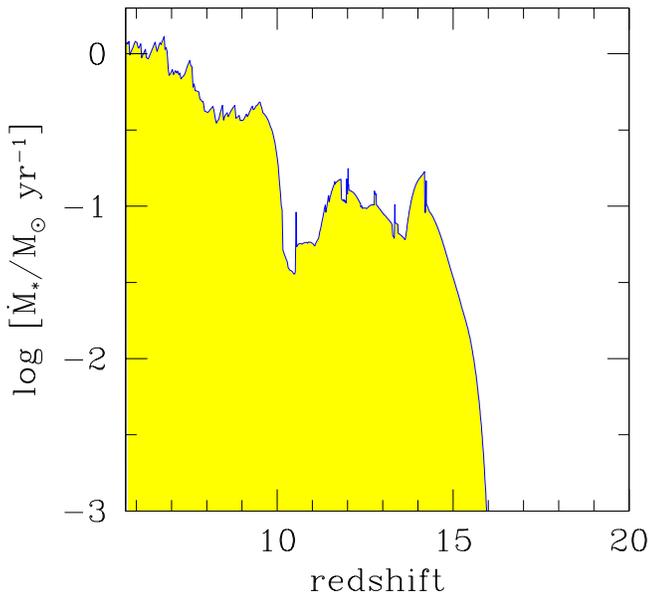}}
  \caption{Star formation history of a typical MW progenitor 
        identified as a LAE at $z\approx 5.7$. }
  \label{SFH}
\end{figure}
\begin{figure}
 \centerline{\psfig{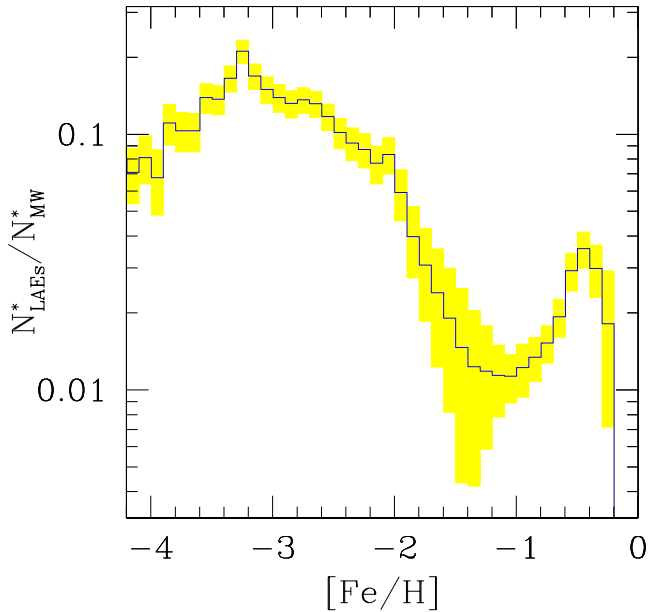}}
 \caption{Fraction of LAE MW halo relic stars as a function of their iron 
abundance. The histogram shows the average fraction among 80 realizations 
of the hierarchical tree; the shaded area shows the $\pm 1\sigma$ dispersion 
among different realizations.}
 \label{mdf}
\end{figure}

As mentioned above, the scatter in the LAE ages originates from different 
assembling and SF histories (SFH) of the MW progenitors. This can be better 
understood by considering the SFH of a typical\footnote{We define as 
``typical'' a LAE whose properties match the average values: 
$M_h\approx 10^{10}\Msun$, $M_g \approx 10^8 \Msun$, 
$ \dot M_* \approx 2.3$~$\Msun/$yr, $t_*\approx 230$~Myr.} MW progenitor 
identified as a LAE, as shown in Fig.~\ref{SFH}. We find that since most 
LAEs typically correspond to the major branch, their progenitor seeds are 
associated with high-$\sigma$ peaks of the density field virializing and 
starting to form stars at high redshifts ($z\approx 16$). The SF rapidly 
changes in time, exhibiting several bursts of different intensities and 
durations, which follow merging events refueling gas for SF. During the 
``quiet'' phase of accretion, instead, SN feedback regulates SF into a more 
gentle regime. The duration and intensity of the peaks depends on the 
effectiveness of these two competitive physical processes. At high redshifts 
($z>10$) the peaks are more pronounced because (i) the frequency of major 
merging is higher and (ii) mechanical feedback is stronger given the shallower 
potential well of the hosting halos ($M\approx 10^{8.5}\Msun$, Salvadori,
Ferrara \& Schneider 2008). 

We finally turn to the last question concerning the contribution of old and 
massive progenitors seen as LAEs, to the very metal-poor stars ([Fe/H]$<-2$) 
observed in the MW halo. In Fig.~\ref{mdf} we show the fractional contribution 
of long-living stars from LAEs to the MDF at $z=0$. LAEs provide $> 10\%$ of 
the very metal-poor stars; the more massive the LAE, the higher is the number 
of [Fe/H]$<-2$ stars they contribute. This is because such metal-poor stellar 
fossils form at $z>6$ in newly virializing halos accreting pre-enriched gas 
out of the MW environment (see Salvadori, Schneider \& Ferrara 2007 and 
Salvadori et~al. 2010). By $z \approx 5.7$, many of these premature building 
blocks have merged into the major branch, i.e. the LAE. Because of the gradual 
enrichment of the MW environment, which reaches [Fe/H]$\approx -2$ at $z\approx 5.7$ 
(see Fig.~1 of Salvadori, Ferrara \& Schneider 2008), most of [Fe/H]$>-2$ stars 
form at lower redshifts, $z < 5.7$, thus producing the drop at [Fe/H]$>-2$. 
Note also the rapid grow of $N^*_{LAEs}/N^*_{MW}$ at [Fe/H]$>-1$, which is 
a consequence of the self-enrichment of building blocks resulting from 
{\it internal} SN explosions. 
\section{Conclusions}
\label{discussion}
We have linked the properties of high-$z$ LAEs to the Local Universe by coupling
the semi-analytical code {\tt GAMETE} to a previously developed LAE model.
According to our results {\it} the progenitors of MW-like galaxies cover a wide 
range of observed Ly$\alpha$ luminosity, $L_{\alpha} =10^{39-43.25}$ erg~s$^{-1}$, 
with $L_{\alpha}$ increasing with $M_*$ (or, equivalently, $M_h$); hence some
of them can be observed as LAEs. In each hierarchical merger history we find 
that, on average, {\it only one} star-forming progenitor (among $\approx 50$) 
is a LAE, usually corresponding to the major branch of the tree ($M_h\approx 
10^{10}\Msun$). Nevertheless, the probability to have {\it at least}
one visible progenitor in any merger history is very high ($P=68\%$). 
Interestingly, we found that the LAE candidates can be also observed as dropout 
galaxies since their UV magnitudes are always $M_{UV} < -18$.

On average the identified LAE stars have intermediate ages, 
$t_* \approx 150-400$~Myr, and metallicities $Z\approx 0.3-1 \Zsun$; 
an exception is represented by five newly formed galaxies, which are hosted 
by small DM halos, $M_h\approx 10^9\Msun$, which are yet visible as (faint) 
LAEs ($L_{\alpha}\approx 10^{42.05}$~erg~s$^{-1}$) by virtue of their high 
star formation rate and extremely young stellar population, $t_* < 5$~Myr. 
The low metallicity of these young galaxies, $Z\approx 0.016-0.044 \Zsun$, 
reflects that of the MW environment at their formation epoch. 
Although rare (5 out of 80 LAEs), these Galactic 
building blocks could be even more unambiguously identified among the least 
luminous LAEs, due to their larger Ly$\alpha$ equivalent widths, 
$EW = 60-130$~\AA\ with respect to those ($\approx 40$~\AA) shown by older 
LAEs\footnote{Due to the low dust mass, $f_c= f_\alpha \approx1$ for all 
the progenitors we identify as LAEs; the observed EW is solely governed by 
$T_\alpha$, which leads $EW \approx 40$ \AA\, for almost all LAEs, except 
those with $t_*\leq 10$ Myr.}. These small and recently virialized halos 
($z_{vir}\lsim 6$) could be the progenitors of Fornax-like dwarf spheroidal 
galaxies (see Fig.~1 of Salvadori \& Ferrara 2009). By identifying these LAEs, 
therefore, it would be possible to observe the most massive dSphs of the MW 
system just at the time of their birth.

Uncertainties remain on the treatment of dust in calculating the LF and 
observed properties of the MW progenitor LAEs identified here, especially 
at the low luminosity end of the LF. Several aspects require additional study. 
As gas, metal and dust are preferentially lost from low mass halos, pushing 
the mass resolution of simulations to even lower masses would be important. 
Also, the amount of dust lost in SN-driven winds remains very poorly understood, 
an uncertainty that propagates in the evaluation of the continuum and Ly$\alpha$ 
radiation escaping from the galaxy. Finally, Ly$\alpha$ photons could be affected 
considerably by the level of dust clumping. It is unclear to what extent these 
effects may impact the visibility of LAEs, as discussed in e.g. Dayal, Ferrara 
\& Saro (2010). Progress on these issues is expected when high-resolution 
FIR/sub-mm observations of LAEs with ALMA will become available in the near 
future (Finkelstein et al. 2009; Dayal, Hirashita \& Ferrara 2010).
\section*{Acknowledgements}
We kindly acknowledge the anonymous referee for his/her positive and useful 
comments. We thank all the DAVID
members\footnote{http://wiki.arcetri.astro.it/bin/view/DAVID/WebHome} for
enlightening discussions, and the Osservatorio Astrofisico di Arcetri for hosting the stimulating DAVID meetings.

\label{lastpage} 
\end{document}